\documentclass[12pt]{article}
\usepackage{amssymb}

\input{tcilatex}
\begin{document}

\bigskip\ 

\bigskip\ 

\begin{center}
\textbf{ORIENTED MATROID THEORY AND LOOP QUANTUM}

\smallskip\ 

\textbf{GRAVITY IN (2+2) AND EIGHT DIMENSIONS}

\textbf{\ }

\textbf{\ }

\smallskip\ 

J. A. Nieto\footnote{%
nieto@uas.uasnet.mx}

\smallskip

\textit{Mathematical, Computational \& Modeling Science Center, Arizona
State University, PO Box 871904, Tempe, AZ 85287, USA}

\textit{Facultad de Ciencias F\'{\i}sico-Matem\'{a}ticas de la Universidad
Aut\'{o}noma de Sinaloa, 80010, Culiac\'{a}n Sinaloa, M\'{e}xico}

\textit{Departamento de Investigaci\'{o}n en F\'{\i}sica de la Universidad
de Sonora, 83000, Hermosillo Sonora , M\'{e}xico}

\bigskip\ 

\bigskip\ 

Abstract

\smallskip\ 
\end{center}

We establish a connection between oriented matroid theory and loop quantum
gravity in (2+2) (two time and two space dimensions) and 8-dimensions. We
start by observing that supersymmetry implies that the structure constants
of the real numbers, complex numbers, quaternions and octonions can be
identified with the chirotope concept. This means, among other things, that
normed divisions algebras, which are only possible in 1,2, 4 or
8-dimensions, are linked to oriented matroid theory. Therefore, we argue
that the possibility for developing loop quantum gravity in 8-dimensions
must be taken as important alternative. Moreover, we show that in
4-dimensions, loop quantum gravity theories in the (1+3) or (0+4) signatures
are not the only possibilities. In fact, we show that loop quantum gravity
associated with the (2+2)-signature may also be an interesting physical
structure.

\bigskip\ 

Keywords: Loop quantum gravity, eight dimensions, (2+2)-dimensions

Pacs numbers: 04.60.-m, 04.65.+e, 11.15.-q, 11.30.Ly

March, 2010

\newpage

It is known that the self-dual (or antiself-dual) concept associated with
the 2-form Riemann tensor $R^{AB}$ plays a central role in quantum gravity 
\textit{a la} Ashtekar [1]-[3]. Mathematically, the self-dual sector of $%
R^{AB}$ is realized by introducing a dual tensor $^{\ast }R^{AB}$ such that
the self-dual curvature,%
\begin{equation}
^{+}R^{AB}=\frac{1}{2}(R^{AB}+\alpha ^{\ast }R^{AB}),  \tag{1}
\end{equation}%
where $\alpha =\{1,i\}$, is again a 2-form. Using the completely
antisymmetric density $\epsilon _{A_{1}..A_{D}}$ ($\epsilon $-symbol) which
takes values in the set $\{-1,0,1\}$ one can define $^{\ast
}R_{A_{1}...A_{D-3}}$ as%
\begin{equation}
^{\ast }R_{A_{1}...A_{D-3}}=\frac{1}{2}\epsilon
_{A_{1}..A_{D-3}A_{D-2}A_{D-1}}R^{A_{D-2}A_{D-1}}.  \tag{2}
\end{equation}%
In this case, one immediately sees that the dual $^{\ast
}R_{A_{0}...A_{D-3}} $ is a 2-form only in $4$-dimensions. This seems to
indicate that, from a quantum gravity perspective, $4$-dimensions is an
exceptional signature. However, in references [4-7] it is shown that also
makes sense to consider self-dual gravity in $8$-dimensions, but one should
define $^{\ast }R^{AB}$ in terms of the $\eta $-symbol.(see Refs. [8-9] and
[10]-[12]) rather that in terms of the $\epsilon $-symbol. In fact, the $%
\eta $-symbol is very similar to the $\epsilon $-symbol in $4$-dimensions;
it is a $4$-index completely antisymmetric object and takes values also in
the set $\{-1,0,1\}$. However, the $\eta $-symbol is defined in $8$%
-dimensions rather than in $4$. Moreover, while the $\epsilon $-symbol in $4$%
-dimensions can be connected with quaternions, the $\eta $-symbol is related
to the octonion structure constants (see Ref. [13] and References therein).
Thus, in $8$-dimensions we can also introduce the dual tensor 
\begin{equation}
^{\ast }R_{A_{1}A_{2}}=\frac{1}{2}\eta _{A_{1}A_{2}A_{3}A_{4}}R^{A_{3}A_{4}},
\tag{3}
\end{equation}%
and consequently the self-dual object $^{+}R^{AB}$, given in (1), is again a 
$2$-form. Since in $2$-dimensions one can always write $R^{AB}=R\epsilon
^{AB}$, $^{\ast }R^{AB}=\quad ^{\ast }R\epsilon ^{AB}$ and $^{+}R^{AB}=\quad
^{+}R\epsilon ^{AB}$, with $R=\frac{1}{2}R^{AB}\epsilon _{AB},$, one sees
that self-duality requirement (1) can also be achieved in $2$-dimensions.
The case of $1$-dimension corresponds to an $\epsilon $ without indices and
may be identified with the real numbers. Therefore, this shows that the set%
\begin{equation}
D=\{1,2,4,8\},  \tag{4}
\end{equation}%
describes the dimensionality of the "spacetime" where self-duality can be
accomplished. One may recognize in (4) the only possible dimensions for a
real division algebras [14]-[15] (see also Ref. [16] and references
therein). Moreover, the set (4) corresponds to the dimensions associated
with the normed division algebras; real numbers, complex numbers,
quaternions and octonions. From the point of view of string theory and
massless vector field, the dimensions in the set $D$ can be understood as
the true physical degrees of freedom, corresponding to dimensions $3,4,6$
and $10$ in the covariant approach, respectively. It turns out interesting
that this normed division algebras are related to the objects $\epsilon $, $%
\epsilon _{AB}$, $\epsilon _{ABCD}$ and $\eta _{ABCD}$, respectively. In
fact, we shall see below that this is not a coincidence but as a result of a
link between supersymmetry, division algebras and oriented matroids.

The next step is to analyze the above scenario of $1,2,4$ or $8$-dimensions
from the point of view of the "spacetime"-signature. The Milnor-Bott [14]
and Kervair [15] theorem for real division algebras and Hurwitz theorem [17]
for normed division algebras refer to Euclidean space, but in Refs [18],
[19] and [20] it is shown that the set $D$ may also be linked to other
signatures. In $2$-dimensions we have the two possible signatures $(1+1)$
and $(0+2)$ which may be identified with $2$-dimensional gravity (see Ref.
[21] and references therein). Traditionally, in $4$-dimensions one assumes
the signatures $(0+4)$ or $(1+3)$, but in this article we will show that the 
$(2+2)$-signature (two time and two space dimensions) may emerge also as
interesting possibility (see Ref. [22] and references therein). Similarly,
in $8$-dimensions one can consider the signatures $(0+8)$, $(1+7)$, $(2+6)$
and $(4+4)$ which in the covariant context may correspond to $(0+10)$, $(1+9)
$, $(2+8)$, $(4+6)$ and $(5+5)$ (see Ref. [23]). Of course, it will be
wonderful to have a theory which predicts no only the dimensionality of the
"spacetime" but also its signature (See Ref. [24]). At least the
self-duality concept predicts the dimensionality of the "spacetime". But in
the lack of a sensible theory which determines the signature of the
"spacetime" we need to explore all possibilities. Eventually this may help
to find, for a fixed dimensionality, a connection between the different
signatures.

Let us analyze the above scenario from the point of view of gauge group
theory. It is known that the algebra $so(1,3)$ can be written as $%
so(1,3)=su(2)\times su(2)$, or the algebra $so(4)$ as $so(4)=so(3)\times
so(3)$, corresponding to the signatures $(1+3)$ and $(0+4)$ respectively.
So, in both cases the curvature $R^{AB}$ can be decomposed additively:%
\begin{equation}
R^{AB}(\omega )=\quad ^{+}R^{AB}(^{+}\omega )+\quad ^{-}R^{AB}(^{-}\omega ),
\tag{5}
\end{equation}%
where $^{+}\omega $ and $^{-}\omega $ are the self-dual and antiself-dual
parts of the spin connection $\omega $. In an Euclidean context, this is
equivalent to write the normed group for quaternions $O(4)$ as $%
O(4)=S^{3}\times S^{3}$, where $S^{3}$ denotes the $3$-sphere. The situation
in $8$-dimensions is very similar since $O(8)=S^{7}\times S^{7}\times G_{2}$%
, with $S^{7}$ denoting the $7$-sphere, suggesting that one can also define
self-duality in $8$-dimensions, but modulo the exceptional group $G_{2}$
[8-9].

In $(2+2)$-dimensions we have analogue situation since $SO(2,2)=$ $%
SU(1,1)\times SU(1,1)$. It is worth mentioning a number of properties of the
group $SU(1,1)$. First of all, the group $SU(1,1)$ is isomorphic to the
groups $SL(2,R)$ and $Sp(2)$. Secondly, just as $SU(2)$ is the double cover
of $SO(3)$, we have that $SU(1,1)$, $SL(2,R)$ and $Sp(2)$ are double cover
of $SO(1,2)$. Moreover, $SU(1,1)$ manifold is topologically $R^{2}\times
S^{1}$. In general, the important role played by the groups $SU(1,1)$, $%
SL(2,R)$ and $Sp(2)$ has been recognized, for long time, in a various
physical scenarios, including $2$-dimensional black-holes [25], 2t physics
[26], and string theory [27]-[28]. In this context, $^{+}\omega $ (or $%
^{-}\omega )$ must be understood as a connection associated with the gauge
groups $SU(1,1)$, $SL(2,R)$ and $Sp(2)$. As a consequence, in $(2+2)$%
-dimensions the self-dual connection $^{+}\omega $ can be linked with the
group $SO(1,2)$. We will see that in this case an interesting possibility
arises at the quantum gravity level.

Let us consider now the Clifford algebra%
\begin{equation}
\Gamma _{\mu }^{AC}\Gamma _{\upsilon CB}+\Gamma _{\nu }^{AC}\Gamma _{\mu
CB}=2\delta _{B}^{A}\eta _{\mu \nu }.  \tag{6}
\end{equation}%
In order to have a supersymmetric Yang-Mills theories it is necessary that $%
\Gamma _{\mu }^{AB}$ satisfies the additional condition (see Ref. [29])

\begin{equation}
\Gamma _{\mu A(B}\Gamma _{CD)}^{\mu }=0,  \tag{7}
\end{equation}%
where the bracket $(BCD)$ means completely antisymmetric. It can be shown
that the two relations (6) and (7) are equivalent to the condition for
normed division algebras. So, the possible dimensions of supersymmetric
Yang-Mills theories are limited to only $1,2,4$ or $8$ (see Refs. [29] and
[20]). The interesting thing that we would like to add is that\ the
expression (7) can be identified with a Grassman-Pl\"{u}cker relation and
consequently the $\Gamma _{\mu }^{AB}$ satisfying (7) is a chirotope [30]
which takes the values $\epsilon $, $\epsilon ^{AB}$, $\epsilon ^{ABCD}$ and 
$\eta ^{ABCD}$ depending if we are considering $1,2,4$ or $8$-dimensions,
respectively. In Refs [31] it is shown that the $\epsilon $-symbol is a
chirotope. Similarly, in connection to maximal supersymmetry in Ref. [32] it
is shown the $\eta $-symbol is also a chirotope. The new ingredient is that
by using the Clifford structure, expressions (6) and (7), both cases, $%
\epsilon $-symbol-chirotope and $\eta $-symbol-chirotope, can also be
obtained. This result suggests a link between maximal supersymmetry and
Clifford structure. Now, there exist a definition of an oriented matroid in
terms of chirotopes [30]. So, we have established a connection between
supersymmetry, division algebras and oriented matroids. It is important to
mention that the set $D=\{1,2,4,8\}$, and the corresponding quantities $%
\epsilon $, $\epsilon ^{AB}$, $\epsilon ^{ABCD}$ and $\eta ^{ABCD}$, can
also connected with the so called $r$-fold cross product [8].

The next step is now to bring these results at the level of canonical
Diffeomorphism and Hamiltonian constraints of quantum gravity. First,
suppose that the Hamiltonian operators $\hat{H}$ and $\hat{H}_{l}$ act on
the physical sates $\mid \Psi >$ in the form

\begin{equation}
\hat{H}\mid \Psi >=0  \tag{8}
\end{equation}%
and

\begin{equation}
\hat{H}_{l}\mid \Psi >=0,  \tag{9}
\end{equation}%
respectively. We shall assume that $\hat{H}$ and $\hat{H}_{l}$ can be
written in terms of the canonical variables $\hat{A}_{i}^{a}$ and $\hat{E}%
_{(a)}^{~~~i}$. Here, $\hat{E}_{(a)}^{~~~i}$ is an operator associated with
the $E_{i}^{~(a)}$ part of the general vielbein on a $M^{D}$-manifold (see
Ref. [7] and references therein)

\begin{equation}
e_{\mu }^{~(A)}=\left( 
\begin{array}{cc}
E_{0}^{~(0)} & E_{0}^{~(a)} \\ 
0 & E_{i}^{~(a)}%
\end{array}%
\right) ,  \tag{10}
\end{equation}%
and $\hat{A}_{i}^{a}$ is an operator associated with the self-dual
connection $^{+}\omega _{i}^{~(0a)}\equiv A_{i}^{a}$.

In the case of $(2+2)$-dimensions one has the constrains

\begin{equation}
H=\frac{1}{4}\tilde{E}\epsilon ^{ijk}~E_{i}^{~(a)}~^{+}R_{jk(0a)}=0  \tag{11}
\end{equation}%
and

\begin{equation}
H_{l}=\frac{1}{4}\tilde{E}\epsilon ^{ijk}~\epsilon
_{~~bc}^{a}E_{i}^{~(b)}E_{l}^{~(c)}~^{+}R_{jk(0a)}=0.  \tag{12}
\end{equation}%
Here, $^{+}R_{jk(0a)}$ is a reduction to seven dimensions of $^{+}R^{AB}$
and $\epsilon ^{ijk}=\frac{1}{\tilde{E}}\varepsilon ^{ijk}$, with $%
\varepsilon ^{123}=1$. Furthermore, $\tilde{E}$ is the determinant of $%
E_{i}^{~~(a)}$. Although these constrains have the same form as the case of $%
(1+3)$-signature there are important differences. First, the symbols $%
\epsilon ^{ijk}$ refers to $(1+2)$-"spacetime" rather than to $(0+3)$.
Second, $^{+}\omega _{i}^{~(0a)}\equiv A_{i}^{a}$ will be $SU(1,1)$ gauge
field rather than $SU(2)$. It turns out useful to change the notation in
(11) and (12) by writing $^{+}R_{jk(0a)}\equiv F_{jka}$, so that $F_{jk}^{a}$
can be identified with the curvature of $A_{i}^{a}$, $F=dA+A\wedge A$. We
also write%
\begin{equation}
P_{a}^{i}=\tilde{E}E_{(a)}^{~i}.  \tag{13}
\end{equation}%
Thus, in terms of $F_{jk}^{a}$ and $P_{i}^{a}$ the constraints (11) and (12)
become (see Ref. [33] and references therein)

\begin{equation}
H=\frac{1}{4\sqrt{\det (P_{a}^{i})}}P_{a}^{i}P_{b}^{j}\epsilon _{\quad
c}^{ab}~F_{ij}^{c}=0  \tag{14}
\end{equation}%
and

\begin{equation}
H_{l}=\frac{1}{2}P_{a}^{i}F_{li}^{a}=0.  \tag{15}
\end{equation}%
Here, we have used the identities $\epsilon ^{ijk}E_{i}^{~(a)}=\epsilon
^{abc}E_{b}^{~~~j}E_{c}^{~~~k}$ and $\epsilon
_{abc}E_{i}^{~(b)}E_{j}^{~(c)}=\epsilon _{ijk}E_{(a)}^{~~~k}$ which can be
derived from $\epsilon ^{ijk}\epsilon
_{abc}E_{i}^{~(a)}E_{j}^{~(b)}E_{k}^{~(c)}=1$.

The only non-vanishing Poisson bracket between the pair of canonical
variables $A_{i}^{a}(x)$ and $P_{a}^{i}(y)$ is

\begin{equation}
\{A_{j}^{a}(x),P_{b}^{i}(y)\}=\delta _{j}^{i}\delta _{b}^{a}\delta (x,y). 
\tag{16}
\end{equation}%
One may assume that the physical states $\mid \Psi >$ can be written in
terms of a Wilson loop wave function

\begin{equation}
\Psi _{\gamma }(A)=trP\exp \int_{\gamma }A,  \tag{17}
\end{equation}%
which satisfies the representation conditions

\begin{equation}
\begin{array}{c}
\hat{A}_{i}^{a}\Psi (A)=A_{i}^{a}\Psi (A), \\ 
\\ 
\hat{P}_{a}^{i}\Psi (A)=\frac{\delta \Psi (A)}{\delta A_{i}^{~a}}.%
\end{array}
\tag{18}
\end{equation}%
Here, the integral (17) is over the loop $\gamma $. If we want to go further
and consider interactions one first needs to make finite computations. The
strategy in this case is to decompose the loop $\gamma \ $in a finite number
of edges $e$, in other words, one represents $\gamma $ as a graph $G$. This
allows us to write the function $\Psi _{\gamma }(A)$ as [1]

\begin{equation}
\Psi _{\gamma }(A)=\psi (h_{e_{1}}(A),...,h_{e_{m}}(A)),  \tag{19}
\end{equation}%
where $h_{e}$ is holonomy along each edge $e$. However, in order to
implement this strategy one needs to complete the computations by
considering all possible graphs $G$. It is worth mentioning that the program
of considering Wilson loops for a gauge field $A$ associated with a
noncompact group $S(1,2)$ has already been considered in the context of $%
(1+2)$-dimensional gravity (see Ref. [25]). Of course, $(2+2)$-dimensional
gravity, with gauge group $SU(1,1)$, is different theory, but at least in
both cases the gauge field $A$ can be associated with the noncompact group $%
SO(1,2)$. One can even think in a connection between the two theories by
assuming a compactification of one of the time dimensions in the $(2+2)$%
-gravitational theory.

In the case of $8$-dimensions, one has that the classical constraints $H$
and $H_{l}$ are given by [7]

\begin{equation}
H=\frac{1}{4}\tilde{E}\eta ^{ijk}~E_{i}^{~(a)}~^{+}R_{jk(0a)}=0  \tag{20}
\end{equation}%
and

\begin{equation}
H_{l}=\frac{1}{4}\tilde{E}\eta ^{ijk}~\eta
_{~~bc}^{a}E_{i}^{~(b)}E_{l}^{~(c)}~^{+}R_{jk(0a)}=0.  \tag{21}
\end{equation}%
It can be expected that these constraints can also be written in the form%
\begin{equation}
H=\frac{1}{4\sqrt{\det (P_{a}^{i})}}P_{a}^{i}P_{b}^{j}\epsilon _{\quad
c}^{ab}~F_{ij}^{c}=0  \tag{22}
\end{equation}%
and

\begin{equation}
H_{l}=\frac{1}{2}P_{a}^{i}F_{li}^{a}=0.  \tag{23}
\end{equation}%
However, one should be careful in this case with the meaning of the
determinant $\det (P_{a}^{i})$ because now it is defined in terms of the
octonion structure constant $\eta ^{ijk}~$and $\eta ^{abc}$ rather than in
terms of $\epsilon ^{ijk}$ and $\epsilon ^{abc}$. In this case one can
choose $A_{i}^{a}$ as a $spin(7)$ gauge field. The formulae (16)-(19) for
the pair of canonical variables $A_{i}^{a}(x)$ and $P_{a}^{i}(y)$ also
applies to this case. This means that in $8$-dimensions one can also use
graph theory to make finite computations. Of course the topology of a given $%
7$-dimensional manifold is more complicated (see Ref. [34]) than in $4$%
-dimensions. Nevertheless, one should expect that a more rich structure may
emerge beyond graph theory. For instance, one may look for physical states
in terms of the analogue of the Chern-Simons states in $4$-dimensions [35].
The reason for this is because Chern-Simons theory is linked to instantons
in $4$-dimensions via the topological term $\int_{M^{4}}tr\epsilon ^{\mu \nu
\alpha \beta }F_{\mu \nu }F_{\alpha \beta }$, while in $8$-dimensions the
topological term should be of the form $\int_{M^{8}}tr\eta ^{\mu \nu \alpha
\beta }F_{\mu \nu }F_{\alpha \beta }$ which can be related with $G_{2}$%
-instantons (see [36] and references therein).

Thus in both cases, in $(2+2)$-dimensions and $8$-dimensions, the loop
quantum gravity approach [37]-[41] indicates that it is necessary for
computations to use directed graph formalism. But a directed graph $G$ is a
particular case of a oriented matroid $\mathcal{M}$. So one may expect that
oriented matroid theory may play an important mathematical tool in new
developments on this program. And, in fact, this seems to have been recently
confirmed [42]. However, we believe that the importance of oriented matroid
theory in loop quantum gravity should be extended beyond graph theory. The
reason for this expectation comes from a number of previous connections
between matroids and different scenarios [43-47], including Chern-Simons
theory , superstrings , p-branes and M-theory. In the process we have even
develop the idea of the gravitoid [42] which refers to any connection
between matroids and gravitons. In all these cases, the main motivation is
the search for a duality principle underlying M-theory. Oriented matroid
theory seems to provide the mathematical tool necessary for this goal, since
one of its central topics is precisely duality. In fact, we have proposed
[43] the oriented matroid theory as the mathematical framework for M-theory.
In the case of loop quantum gravity similar duality motivation can be
considered. This idea emerges natural since we have proved that in $(2+2)$
and $8$-dimensions, oriented matroid theory is linked to loop quantum
gravity at both levels, namely the constraints operators (Heisenberg-like
approach) and physical states (Schr\"{o}dinger-like approach). Since one can
associate with every oriented matroid $\mathcal{M}$ a dual matroid $\mathcal{%
M}^{\ast }$. One should expect that duality also plays a central role in
loop quantum gravity. Let us outline how this can be accomplish. The
following arguments are, in fact, true for any of the dimensions $1,2,3$ or $%
8$ and any of the corresponding signatures.

We shall be brief in our comments (see the Ref. [48] for details). Consider
any graph $G$. Let $B$ be the incidence matrix of $G$. One can introduce a
pair of complementary subspaces $L,L^{\perp }$ in $R^{m}$, where $m$ is the
number of edges in $G$, which can be can be associated with $B$ by the
expressions $L=\ker B$ and $L^{\perp }=imB$. Indeed, $L$ and $L^{\perp }$
corresponds to the circuit and cocircuit space of $G$. It turns out that $L$
and $L^{\perp }$ satisfy the so called Farkas property: For every edge $e$
in $G$ either

a) $\exists X\in L,$ $e\in $supp$X,X\geqslant 0$

\noindent or

b)$\exists Y\in L^{\perp },$ $e\in $supp$Y,Y\geqslant 0$

\noindent but not both.

\noindent Here, $X$ and $Y$ are the incidence vectors associated with a
circuit and cocircuit respectively. Note that this property is self-dual in
the sense that both alternatives a) and b) can be interchanging by replacing 
$L$ by $L^{\perp }$. The central idea in oriented matroid theory is to
generalize this property to any pair of signed sets $(S,S^{\prime })$, with $%
S^{\prime }$ properly defined, such that $(S,S^{\prime })$ satisfies the
analogue of the Farkas property. In fact, an oriented matroid can be defined
in terms of the pair $(S,S^{\prime })$ and such a generalized Farkas
property. One interesting thing is that given this definition of an oriented
matroid one finds that there are oriented matroids which can not be realized
as graphs. So the oriented matroid notion is a more general structure than
the graph concept. Another interesting aspect of this construction of
oriented matroids is that the two spaces $L$ and $L^{\perp }$ (or $S$ and $%
S^{\prime })$ are equally important. This is one of the reasons why every
oriented matroid $\mathcal{M}$ has always a dual $\mathcal{M}^{\ast }$.

How this definition of an oriented matroid in terms of the Farkas property
can be linked to loop quantum gravity? Let us assume that a physical state
has the form

\begin{equation}
\Psi _{C}(A,L)=trP\exp \int_{C}A,  \tag{24}
\end{equation}%
where $C$ is a circuit of a given graph $G$. We write $\Psi _{C}(A,L)$ to
emphasis that $C$ is contained in the circuit space$\ L=\ker B$, with $B$
the incidence matrix of $G$. But according to the Farkas property it must be
equally important to consider the physical state,

\begin{equation}
\Psi _{C^{\ast }}(A^{\ast },L^{\perp })=trP\exp \int_{C^{\ast }}A^{\ast }. 
\tag{25}
\end{equation}%
Here, $C^{\ast }$ is a cocircuit in $L^{\perp }$ and $A^{\ast }$ is a dual
gauge field. Observe that (25) completely dualize (24). This Schr\"{o}%
dinger-like schema for $\Psi _{C}(A,L)$ and $\Psi _{C^{\ast }}(A^{\ast
},L^{\perp })$ must have Heisenberg-like counterpart in terms of dual
Hamiltonian operators constraints. In principle these dual Hamiltonian
constraints can be $\hat{H}$ and $\hat{H}_{l}$ themselves. However, in a
more general scenario one must consider dual Hamiltonian operators
constrains $\hat{H}^{\ast }$ and $\hat{H}_{l}^{\ast }$ acting on the
physical states $\mid \Psi ^{\ast }>$ associated with $\Psi _{C^{\ast
}}(A^{\ast },L^{\perp })$. In other words one must have the symbolic formulae

\begin{equation}
\hat{H}^{\ast }\mid \Psi ^{\ast }>=0  \tag{26}
\end{equation}%
and

\begin{equation}
\hat{H}_{l}^{\ast }\mid \Psi ^{\ast }>=0.  \tag{27}
\end{equation}%
Going backwards the constraints operators $\hat{H}^{\ast }$ and $\hat{H}%
_{l}^{\ast }$ must come from classical constrains $H^{\ast }$ and $%
H_{l}^{\ast }$ which in turn should be possible to derive from a dual
gravitational field $E^{\ast }$ and dual connection $\omega ^{\ast }$ via
the corresponding self-dual curvature $^{+}R^{\ast AB}$. Note that we have
distinguish between two different dualities in $^{+}R^{\ast AB}$. This is
because we are considering the most general dual theory but at some level
one should expect that both kind of dualities are related. In S-duality for
linearized gravity [49], for instance, one starts with a curvature $%
^{+}R^{AB}$ and finds the dual curvature $^{+}W^{AB}$ which can be
identified with $^{+}R^{\ast AB}$. Some of these ideas are under intensive
research and we expect to report our results elsewhere.

It is worth mentioning that a possible connection between oriented matroid
theory and Ashtekar formalism is mentioned in the Refs. [4], [5], [6], [7],
[40] and [42]. Further, in the literature (see [50]-[51] and references
therein) exist a canonical approach of the $(2+2)$-imbedding, but this
should be called $((1+1)+(0+2))$-imbedding since refers to the $(1+3)$%
-signature rather than to the case of $2$-time and $2$-space dimensions ($%
(2+2)$-dimensions) which we have been considered in this work. Nevertheless,
it may be interesting for further research to see if there is a link between 
$((1+1)+(0+2))$-imbedding and $(2+2)$-loop quantum gravity.

Perhaps, the link between oriented matroid theory and loop quantum gravity
may provide new fascinating insights into other contexts in which $(2+2)$%
-signature makes its appearance, including qubit-strings [52] and N=2
strings [53].

\bigskip\ 

\noindent \textbf{Acknowledgments: }I would like to thank M. C. Mar\'{\i}n,
R. Perez-Enriquez and A. Le\'{o}n for helpful comments and to the
Mathematical, Computational and Modeling Science Center at the Arizona State
University for the hospitality, where part of this work was developed.

\smallskip\

\end{document}